\documentclass{JHEP3}
\preprint{LU-TP 02-34}

\usepackage{epsfig}
\usepackage{axodraw}
\usepackage{inputenc}
\usepackage{xspace}
\inputencoding{latin1}
 \renewcommand\email[1]{{\scriptsize\tt\href{mailto:#1}{#1}}}

\newcommand{\as}{\ensuremath{{\alpha}_{s}}}

\newcommand{\kT}{\ensuremath{k_{\perp}}}
\newcommand{\qT}{\ensuremath{q_{\perp}}}
\newcommand{\pT}{\ensuremath{p_{\perp}}}
\newcommand{\ET}{\ensuremath{E_{\perp}}}
\newcommand{\ETmax}{\ensuremath{E_{\perp\max}}}

\newcommand{\kTpot}[1]{\ensuremath{k_{\perp}^{#1}}}
\newcommand{\qbar}{\ensuremath{\overline{q}}}

\newcommand{\cascade}{C\scalebox{0.8}{ASCADE}\xspace}
\newcommand{\ldcmc}{\scalebox{0.8}{LDCMC}\xspace}
\newcommand{\pythia}{P\scalebox{0.8}{YTHIA}\xspace}
\newcommand{\herwig}{\scalebox{0.8}{HERWIG}\xspace}
\newcommand{\pycell}{\scalebox{0.8}{PYCELL}\xspace}
\newcommand{\jimmy}{J\scalebox{0.8}{IMMY}\xspace}
\newcommand{\isajet}{I\scalebox{0.8}{SAJET}\xspace}
\newcommand{\ariadne}{A\scalebox{0.8}{RIADNE}\xspace}
\newcommand{\ftwo}{\ensuremath{F_2}}
\def\mrm#1{\mathrm{#1}}
\def\sub#1{\ensuremath{_{\mrm{#1}}}}

\newcommand{\LDCz}{{\boldmath\bf LDC$_0$}\xspace}
\newcommand{\LDCzp}{{\boldmath\bf LDC$'_0$}\xspace}
\newcommand{\LDCP}{{\boldmath\bf LDC$_P$}\xspace}

\newcommand{\LDCG}{{\boldmath\bf LDC$_G$}\xspace}
\newcommand{\LDCGp}{{\boldmath\bf LDC$'_G$}\xspace}
\newcommand{\PYTz}{{\boldmath\bf PYT$_0$}\xspace}

\newcommand{\PYTS}{{\boldmath\bf PYT$_3$}\xspace}
\newcommand{\PYTD}{{\boldmath\bf PYT$_4$}\xspace}
\newcommand{\PYTP}{{\boldmath\bf PYT$_1$}\xspace}
\newcommand{\PYTDp}{{\boldmath\bf PYT$'_{4}$}\xspace}

\skip\footins = 1\bigskipamount plus 2pt minus 4pt

\title{\boldmath Hadronic Collisions in the\\ Linked Dipole Chain Model}

\author{G\"{o}sta Gustafson, Leif L\"onnblad and Gabriela Miu\\
  Dept.~of Theoretical Physics,
  S\"olvegatan 14A, S-223 62  Lund, Sweden\\
  E-mail: \email{Gosta.Gustafson@thep.lu.se},
  \email{Leif.Lonnblad@thep.lu.se} and
  \email{Gabriela.Miu@thep.lu.se}}
  
\abstract{We apply the Linked Dipole Chain model to hadronic
  collisions using a modified version of the \ldcmc Monte Carlo
  program. In particular we investigate the effects of multiple
  scatterings, which in this framework are reformulated in terms of
  the production of multiple chains of dipoles. We find, among other
  things, that this way of describing multiple scatterings is less
  sensitive to the treatment of soft, non-perturbative, scatterings.
    
  Although the results presented here are only at parton level, by
  comparing with corresponding results from \pythia, we are confident
  that our model will be able to describe essential features of high
  energy hadronic collisions, including underlying events and
  jet-pedestal effects in high $\pT$ events. The model should also be
  applicable to estimates of transverse energy flow in high energy
  heavy ion collisions, of interest in analyses of signals for
  quark-gluon plasma formation.}

\begin{document}
 
\sloppy

\section{Introduction}
\label{sect-intro}

Although the main aim for the future Large Hadron Collider at CERN is
to discover and study the Higgs boson, and to possibly discover new
physics, it is important to keep in mind that our understanding of the
strong interactions, which make up the bulk of the cross section in
hadronic collisions, is far from complete.  The hard production of
large transverse momentum jets may be described to a reasonable
precision by perturbative QCD, but for the dominating soft and
semi-hard interactions we have to rely on phenomenological models.
There may be some gold-plated signals for new physics, where these
so-called minimum bias interactions may be disregarded, but as soon as
observables include jets, a more detailed understanding of the event
structure is needed. Not only is the luminosity so high at the LHC
that any signal event will be accompanied by several overlayed minimum
bias collisions. In addition the inclusive cross section for
parton-parton scatterings becomes larger than the total
non-diffractive cross section already at present day energies, which
implies that there are several parton-parton scatterings in each
single hadron-hadron collision.  Such additional scatterings can be
rather soft, and can therefore not be described by purely perturbative
calculations. The production of minijets is also of interest in high
energy heavy ion collisions, and the analysis of signals for plasma
formation.  They are essential for the initial conditions in studies
of the flow in either a quark-gluon plasma or a hadronic phase
\cite{Huovinen:1998tq}.

Further studies of hadronic collisions are also interesting in its own
right, in order to understand the nature of strong interaction and the
interplay between soft and hard physics.  Many models have been
presented to describe and simulate hadronic collisions.  The Fritiof
model \cite{Andersson:1987gw} successfully describes hadron collisions
at lower energies, up to the ISR range. It is based on a soft momentum
transfer and gluon bremsstrahlung due to the separation of colour
charges. These charges are in the model connected by strings of the
same nature as in $e^{+}e^{-}$-annihilation. For higher energies hard
sub-collisions were added~\cite{Andersson:1993iq}. The formalism did,
however, not give a unique description for how to combine the soft and
hard subprocesses, and therefore the predictive power at high energies
is reduced.
 
The Dual Parton Model (DPM) \cite{Capella:1981xr,Capella:1994yb} aims
at a combined description of soft and hard interactions in an
eikonalized form, including besides a soft and a hard pomeron also
triple- and loop-pomeron contributions. The aim is to describe not
only non-diffractive events, but also elastic scattering and single
and double diffraction, and the model is able to reproduce a large set
of experimental data \cite{Aurenche:1992pk}.
 
At high energies, in the $\bar{p}p$ collider regime, the cross section
for hard or semi-hard parton sub-collisions becomes large, and the
properties of non-diffractive events appear to be dominated by these
subprocesses. In this energy range it is therefore essential to have a
good description of jets from hard parton collisions, including
initial and final state bremsstrahlung.  In \isajet \cite{Paige:1998ux}
and \herwig \cite{Corcella:2000bw} a description of the hard
sub-collisions is complemented by an "underlying event" due to
interactions between the beam remnants. In \pythia
\cite{Sjostrand:2000wi} this underlying event is modeled by including
multiple parton interactions in the generation of the hard parton
collision, rather than adding them as an independent component to the
hard sub-collisions.

Even if the parton-parton collisions are of perturbative nature, with
a \pT\ above a couple of GeV, there are many difficulties, which
cannot be solved by perturbative methods alone. The parton
sub-collisions are typically modeled using collinear factorization.
This gives a divergent expression for the inclusive cross section at
small $p_\perp$, which has to be treated by some kind of cutoff.
Clearly collisions with large impact parameter must be suppressed due
to screening effects, when the impact parameter is larger than the
proton radius. The impact parameter dependence is also expected to
influence the correlations between sub-collisions in a single event.
It is natural to assume that central collisions have a higher activity
than more peripheral collisions and therefore, on average, a larger
number of sub-collisions \cite{Sjostrand:1987su}. Both these effects,
the \pT\ cutoff and the multiplicity correlations, have to be adjusted
to experimental data, which makes it more difficult to make reliable
extrapolations to higher energies.

In collinear factorization the hard partonic cross section is
convoluted with parton densities, which are evolved using DGLAP
\cite{Gribov:1972ri,Lipatov:1975qm,Altarelli:1977zs,Dokshitzer:1977sg}
evolution. The parton densities are fitted to data on e.g.\ \ftwo\ in
DIS and high \ET\ jets in hadronic collisions. For moderate \pT\ at
very large c.m.\ energies, $\sqrt{s}$, the parton evolution contains
large logarithms of $s/\pT^2$ which need to be resummed to all orders.
For these \pT\-values it would therefore be more appropriate to use
BFKL evolution \cite{Kuraev:1977fs,Balitsky:1978ic}, which gives the
correct description, at least for asymptotic energies and fixed
coupling.  The BFKL evolution is related to $k_\perp$-factorization,
and in ref.~\cite{Gustafson:1999kh} it is argued that this approach
gives a dynamical suppression for small \pT, which reduces the
infrared sensitivity.  This is mainly a consequence of a suppression
of the partonic cross section when the virtualities of the colliding
partons are larger than the transverse momentum exchanged in the hard
collision.

The study in ref.~\cite{Gustafson:1999kh} is based on the Linked
Dipole Chain (LDC) model \cite{Andersson:1996ju,Andersson:1998bx} for
DIS. This model is based on the observation that the dominant features
of the parton evolution is determined by a subset of emitted gluons,
which are ordered in both positive and negative light-cone components.
These \emph{primary} gluons make up a linked chain of colour dipoles,
and the sum over all such possible chains, convoluted with
non-perturbative input parton densities, will give the proton
structure function. (For exclusive properties of the final states also
final state radiation has to be included.)  The LDC model is
completely forward-backward symmetric -- i.e.\ it does not matter if
the evolution is performed from the proton towards the virtual photon
or vice versa. This means that LDC is also very suited for a
description of hadronic collisions, where the sum over all possible
chains, convoluted with input parton densities from both hadrons, will
give the inclusive partonic cross section.  The fact that transverse
momenta can go up and down in a single parton chain implies that such
a chain can correspond to more than one hard parton-parton
sub-collision. Thus one type of correlations between the
sub-collisions is automatically included in the LDC formalism.
However, even so the \textit{per chain} cross section exceeds the
total non-diffractive cross section. Therefore there will in general
be several dipole chains stretched between two colliding hadrons, and
it is also in this approach necessary to make assumptions about the
impact parameter dependence, now for the distribution in the number of
chains in one event.

In this paper we will investigate the effects of such multiple chains
in more detail, using the LDC event generator \ldcmc
\cite{Kharraziha:1998dn}, modified to generate hadronic collisions.
The investigation will still be on a somewhat qualitative level since
effects of colour connection between the chains, and of hadronization
are not included.  Therefore we will here compare our results with the
\pythia generator on parton level rather than with experimental data.

A very essential result from these MC studies is the fact that the
number of chains per event is stable against variations of the \pT\ 
cutoff in the perturbative parton evolution. When fitting experimental
data on $F_2$, the input soft gluon distribution is adjusted, and
corresponds to a scale given by the cutoff. The soft input gluons can
also form direct chains between the colliding hadrons, and these
chains are also allowed to emit final state bremsstrahlung below the
cutoff.  If the soft cutoff is increased there will be fewer hard
chains, which is compensated by a more singular input gluon
distribution. As we will show below, when the cutoff is increased, the
decrease in the number of hard chains is compensated by a larger
number of soft chains (which are now also allowed to radiate more).
Thus the total number of chains in $pp$ collisions can be fully
determined by data on $F_2$, without any additional adjustable free
parameter.

The outline of this paper is as follows. In section \ref{sect-DIS} we
give some further comments on parton evolution in DIS, including a
brief introduction to the LDC model.  In section \ref{sect-LDC} we
discuss the application of the LDC model for hadronic collisions. Then
in section \ref{sect-MI} we discuss some general aspects of multiple
scatterings, in particular how they are included in the \pythia
generator and how we treat them in the present analysis. The technical
details of the Monte-Carlo implementation are discussed in section
\ref{sec-mc}. In section \ref{sect-results} we present some
preliminary results, followed by conclusions in section
\ref{sect-conclusions}.

\section{Small-$x$ physics in DIS}
\label{sect-DIS}

\FIGURE[t]{
\begin{picture}(170,230)(0,0)
\ArrowLine(10,15)(50,15)
\Text(20,25)[]{\large $proton$}
\Line(50,20)(100,20)
\Line(50,15)(100,15)
\Line(50,10)(100,10)
\Line(50,20)(80,40)
\GOval(50,15)(10,7)(0){1}
\Text(60,35)[]{\large $k_{0}$}
\Text(130,40)[]{\large $q_{1}$}
\Line(80,40)(120,40)
\Line(80,40)(95,70)
\Text(80,55)[]{\large $k_{1}$}
\Text(145,70)[]{\large $q_{2}$}
\DashLine(87.5,55)(110,55){2}
\Text(90,85)[]{\large $k_{2}$}
\Text(155,100)[]{\large $q_{3}$}
\Line(95,70)(135,70)
\Line(95,70)(105,100)
\DashLine(100,70)(120,80){2}
\Line(105,100)(145,100)
\Line(105,100)(110,130)
\Line(110,130)(150,130)
\Line(110,130)(110,160)
\Line(110,160)(150,160)
\DashLine(120,160)(140,150){2}
\DashLine(130,155)(145,155){2}
\ArrowLine(20,220)(70,200)
\Text(166,160)[]{\large $q_{n+1}$}
\Text(100,145)[]{\large $k_{n}$}
\Text(20,210)[]{\large $lepton$}
\Text(100,185)[]{\large $q_{\gamma}$}
\ArrowLine(70,200)(120,220)
\Photon(70,200)(110,160){3}{5}
\end{picture}  

  \caption{\label{fig:fanDIS} A fan diagram for a DIS event. 
    The quasi-real partons from the initial-state radiation are
    denoted $q_i$, and the virtual propagators $k_i$. The dashed lines
    denote final-state radiation.}}

Effects of BFKL evolution have been studied extensively at HERA, where
BFKL predicted a strong power-like increase of \ftwo\ for small $x$.
It turns out, however, that BFKL suffers from huge next-to-leading
order corrections \cite{Fadin:1998py} and it has been very difficult
to obtain reliable predictions, especially when it comes to
non-inclusive observables involving final-state hadrons. More reliable
predictions for final-state observables can be obtained using CCFM
evolution \cite{Ciafaloni:1988ur,Catani:1990yc}, which agrees with
BFKL in the asymptotic limit, but also approximates DGLAP for large
$x$.  The central feature of the CCFM model is the angular ordering of
the parton chains contributing to the proton structure function, which
is a consequence of colour coherence or soft gluon interaction.  The
parton chains are ordered in energy (or positive light-cone momentum
$q_+$) and in angle (or rapidity $y=\frac{1}{2}\ln(q_+/q_-)$).  (We
use $q_i$ and $k_i$ to denote the real emitted partons and the virtual
links respectively, as indicated in figure \ref{fig:fanDIS}.)  Other
kinematically allowed emissions (symbolized by the dashed lines in
figure \ref{fig:fanDIS}) are treated as final-state emissions.

\subsection{CCFM Evolution}

Like BFKL evolution, the CCFM model is based on the
$\kT$-factorization formalism, and the unintegrated distribution
function is given by
\begin{eqnarray}
  \label{fCCFM}
  \mathcal{G}(x,k_\perp^2,\qbar) &\sim& \sum_n \int \prod^n \bar{\alpha} 
  dz_i \frac{d^2 q_{\perp i}}{\pi q_{\perp i}^2} 
  \left(\frac{1}{z_i}\Delta_{ne}(z_i, k_{\perp i}^2, \qbar_i) +
    \frac{1}{1-z_i}\right) \Delta_S \times\\
  & & \delta(x-\Pi z_i)
  \theta(\qbar_i - \qbar_{i-1} z_{i-1})
  \delta(k_\perp^2-k_{\perp n}^2) \theta(\qbar-\qbar_n z_n)\nonumber
\end{eqnarray}
where the non-eikonal and the Sudakov form factors are given by
\begin{equation}
  \Delta_{ne} = \exp\left(-\bar{\alpha} \ln\frac{1}{z}
    \ln\frac{\kT^2}{z \bar{q}^2}\right) ;
  \,\,\,\,\Delta_S = \exp\left(-\bar{\alpha} \int 
    \frac {d q_{\perp i}^2}{q_{\perp i}^2}
    \frac {dz}{1-z} \Theta_{order}\right)
  \label{formfactors}
\end{equation}
Here $\bar{\alpha} \equiv 3 \alpha_s/\pi$, the splitting parameter $z$
is defined as $z_i=k_{+,i}/k_{+,i-1}$, and the kinematic range allowed
by the ordering constraints is specified by $\Theta_{order}$. The
quantity $\qbar_i$ is defined by $\qbar_i \equiv q_{\perp i}/(1-z_i)$,
which implies that the angular ordering condition is satisfied by the
constraint $\qbar_i > \qbar_{i-1} z_ {i-1}$, accounted for by a
$\theta$-function in eq.~(\ref{fCCFM}).  The CCFM model has been
developed assuming purely gluonic chains, and only the singular terms
in the splitting functions, proportional to $1/z$ and $1/(1-z)$, are
included in a fully consistent way. The $1/z$ pole is most important
in the BFKL region for small $x$, while the $1/(1-z)$ pole and the
Sudakov form factor are essential for larger $x$ and large $\kT$.  We
also note that the distribution function, $\mathcal{G}$, depends on
two separate scales. Besides the transverse momentum, \kT, of the
interacting gluon, it also depends on \qbar, which determines an angle
beyond which there is no (quasi-) real parton in the chain.

A Monte Carlo implementation of CCFM, \cascade \cite{Jung:2001hx},
reproduces a wide range of final-state observables at HERA, although
there are still some unsolved questions. Since CCFM does not include
the non-singular terms in the gluon splitting function, DGLAP is not
fully reproduced in the relevant limit, and including non-singular
terms degrades the reproduction of HERA data, especially for
observables which are supposed to be sensitive to BFKL effects, such
as forward jet cross sections \cite{Anderson:2002cf}.

\subsection{The Linked Dipole Chain Model}

The Linked Dipole Chain (LDC) model
\cite{Andersson:1996ju,Andersson:1998bx} is a reformulation and
generalization of CCFM. It is based on the observation that the
dominant features of the parton evolution is determined by a subset of
emitted gluons, which are ordered in both positive and negative
light-cone components, and also satisfies the relation
\begin{equation}
q_{\perp i} > \min(k_{\perp i},k_{\perp,i-1}).
\end{equation}
\label{eq:ldccut1}
In LDC this subset of ``\emph{primary}'' gluons forms a chain of
initial state bremsstrahlung (ISB), and all other emissions are
treated as final state bremsstrahlung (FSB).

This redefinition of the ISB--FSB separation implies that one single
chain in the LDC model corresponds to a set of CCFM chains.  As was
shown in ref.~\cite{Andersson:1996ju}, when one considers the
contributions from all chains in this set, with their corresponding
non-eikonal form factors, they just add up to one. Thus, the
non-eikonal form factors do not appear explicitly in LDC, and the
gluon distribution function is given by
\begin{eqnarray}
  \label{fLDC}
  \mathcal{G}(x,k_\perp^2) &\sim& \sum_n \int \prod^n \bar{\alpha} 
  dz_i \frac{d^2 q_{\perp i}}{\pi q_{\perp i}^2} 
  P(z_i) \Delta_S \times\\
  & & \theta(q_{+,i-1} -q_{+ i}) \theta(q_{- i} -q_{-,i-1}) 
  \delta(x-\Pi z_i) \delta(\ln k_\perp^2 - \ln k_{\perp n}^2).\nonumber
\end{eqnarray}
Here $P(z)$ is the full gluonic splitting function, and $\Delta_S$ the
associated Sudakov form factor, defined by
\begin{equation}
  \Delta_S = \exp\left(-\bar{\alpha} \int
    \frac {d q_{\perp i}^2}{q_{\perp i}^2}
    z dz P(z) \Theta_{order}\right)
\label{SudLDC}
\end{equation}
where again $\Theta_{order}$ specifies the region of phase space
allowed by the ordering constraints.  The result of this reformulation
is a simpler form for the unintegrated distribution functions, which
essentially depends only on a single scale, $\kTpot{2}$.  As many of
the gluons which make up the initial-radiation chain in the CCFM model
are treated as final state radiation in the LDC formalism, most of the
problem of angular ordering is postponed to the treatment of the final
state radiation.  The ordering of the LDC evolution in both $q_+$ and
$q_-$, and the fact that the $1/z$ pole in the splitting function is
here not associated with an non-eikonal form factor, imply that this
contribution is completely forward--backward symmetric.  Therefore the
chain in figure \ref{fig:fanDIS} can be thought of either as an
evolution from the proton towards the photon or, equivalently, from
the photon towards the proton end. The $1/(1-z)$ pole in $P(z)$ and
the Sudakov form factor are most essential in the DGLAP region, where
chains with ordered $\qT$ values dominate and large $z$-values are
important.  In the LDC model the expression in eq.~(\ref{fLDC}) is
symmetrized so that steps downwards in transverse momentum with large
values of the backwards splitting parameter $z_{-i} =
k_{-i}/k_{-,i+1}$ are weighted by the splitting function $P(z_{-i})$
and a correspondingly defined Sudakov form factor. The absence of the
non-eikonal form factor also implies that quark links and non-singular
terms in the splitting function can be included in a natural and
straightforward way.

The set of ``\emph{primary}'' gluons in eq.(\ref{fLDC}) make up a
chain of linked colour dipoles, and the sum over all such possible
chains, convoluted with non-perturbative input parton densities, will
give the proton structure function.  Besides this inclusive
description of the events, the result in eq.~(\ref{fLDC}) can also be
interpreted as the production probability for an exclusive final
state.  We note, however, that due to the different separation between
ISB and FSB in LDC and CCFM or other approaches, it is important to
include final state emissions before comparing results for exclusive
final states with results from other formalisms or with experimental
data.

The LDC model is also implemented in a MC event generator, \ldcmc.
When the soft input parton densities are adjusted to fit data on
$F_2$, the model is able to reproduce a wide range of HERA data
\cite{Kharraziha:1998dn}. The resulting integrated gluon distribution
also agrees well with global fits by the CTEQ \cite{Lai:1999wy} and
MRST \cite{Martin:2001es} collaborations \cite{Gustafson:2002jy}.
However, as for the CCFM Monte Carlos, the data for forward jets can
only be described if non-singular terms are omitted from the gluon
splitting function.

The fact that the LDC formalism is manifestly forward-backward
symmetric implies that it automatically takes into account
contributions from ``resolved virtual photons'', where the chain
describes evolution from the proton and photon ends towards a central
hard parton sub-collision. This property is a consequence of the
specific choice of separation between initial and final state
emissions adopted in the LDC model, and it means that the formalism is
also very well suited to describe hard interaction in hadron-hadron
collisions, as will be discussed in more detail in the following
section.

\section{The Linked Dipole Chain Model for Hadronic Collisions}
\label{sect-LDC}

As mentioned in section \ref{sect-DIS}, to get the gluon distribution
function in a proton, the perturbative chains in eq.~(\ref{fLDC}) have
to be convoluted with non-perturbative input parton densities,
$g_0(x_0,k_{\perp 0}^2)$, for $k_{\perp}^2 = k_{\perp 0}^2$:
\begin{equation}
  g(x,k_{\perp}^2) = \int dx_0 dx' \mathcal{G}(x',\kT^2;k_{\perp 0}^2) 
  g_0(x_0,k_{\perp 0}^2) \delta(x_0 x' - x)
\label{convol}
\end{equation}
As we are assuming a running coupling, which diverges for small $\qT$,
it is necessary to introduce a cutoff, $k_{\perp 0}$, for the
$q_{\perp i}$ distributions in eq.~(\ref{fLDC}) (see e.g.\ 
\cite{Andersson:1998bx}). In eq.~(\ref{convol}) this is explicitely
indicated as an argument in $\mathcal{G}$.  Thus $g_0(x_0,k_{\perp
  0}^2)$ is assumed to describe all chains with transverse momenta
below the cutoff $k_{\perp 0}$.
 
\FIGURE[t]{\scalebox{0.8}{\mbox{
\begin{picture}(140,230)(-10,-50)
  \Text(-10,15)[]{\large $P_a$}
  \Line(0,15)(40,15)
  \Line(40,20)(70,20)
  \Line(40,15)(70,15)
  \Line(40,10)(70,10)
  \GOval(40,15)(10,7)(0){1}
  \Line(50,20)(60,40)
  \Text(47,34)[]{\large $k_{0}$}
  \Line(60,40)(90,40)
  \Text(100,40)[]{\large $q_{1}$}
  \Line(60,40)(70,70)
  \Text(55,55)[]{\large $k_{1}$}
  \Line(70,70)(105,70)
  \Text(115,70)[]{\large $q_{2}$}
  \Line(70,70)(75,100)
  \Text(65,85)[]{\large $k_{2}$}
  \Line(75,100)(110,100)
  \Text(120,100)[]{\large $q_{3}$}
  \DashLine(75,100)(75,130){2}
  \DashLine(75,130)(70,160){2}
  \Line(70,160)(105,160)
  \Text(115,160)[]{\large $q_{n-1}$}
  \Line(70,160)(60,190)
  \Text(55,173)[]{\large $k_{n-1}$}
  \Line(60,190)(90,190)
  \Text(100,190)[]{\large $q_{n}$}
  \Line(60,190)(50,210) 
  \Text(47,196)[]{\large $k_{n}$}
  \Text(-10,215)[]{\large $P_b$}
  \Line(0,215)(40,215)
  \Line(40,220)(70,220)
  \Line(40,215)(70,215)
  \Line(40,210)(70,210)
  \GOval(40,215)(10,7)(0){1}
  \Text(40,-20)[]{\large (a)}
\end{picture}
}} 
\hspace{3cm}\scalebox{0.8}{\mbox{
\begin{picture}(140,230)(-10,-50)
  \Text(-10,15)[]{\large $P_a$}
  \Line(0,15)(40,15)
  \Line(40,20)(70,20)
  \Line(40,15)(70,15)
  \Line(40,10)(70,10)
  \GOval(40,15)(10,7)(0){1}
  \Text(40,115)[]{\large $k_{0}$}
  \CArc(-489,115)(547,-10,10)
  \Text(-10,215)[]{\large $P_b$}
  \Line(0,215)(40,215)
  \Line(40,220)(70,220)
  \Line(40,215)(70,215)
  \Line(40,210)(70,210)
  \GOval(40,215)(10,7)(0){1}
  \Text(40,-20)[]{\large (b)}  
\end{picture}}}

  \caption{\label{fig:ppchain} (a) A chain of $n$ initial state
    emissions in a hadronic collision. (b) An exchange of a single
    soft gluon without initial state emissions.}}

\subsection{Hard Chains}

The fact that $\mathcal{G}$ is completely forward-backward symmetric,
as discussed in section \ref{sect-DIS}, now implies that the cross
section, $\sigma_c$, for formation of a chain between two colliding
protons is directly obtained if the distribution in eq.~(\ref{convol})
is convoluted with a parton distribution from the upper proton in
figure \ref{fig:ppchain}.  Thus we find:
\begin{equation}
  \sigma_c = \frac{\pi^3}{2 k_{\perp 0}^2} \int dx_{0+} dx\, dx_{0-} 
  g_0(x_{0+},k_{\perp 0}^2) 
  \mathcal{G}(x,k_{\perp 0}^2;k_{\perp 0}^2) g_0(x_{0-},k_{\perp 0}^2) 
  \delta(x_{0+} x\, x_{0-}-k_{\perp 0}^2/s),
\label{sigmac}
\end{equation}
with $x_{0+}=k_{0+}/P_{a+}$ and $x_{0-}=k_{n-}/P_{b-}$. The
normalization factor $\pi^3/2$ originates from the fact that a hard
scattering in the middle of the chain, with $q_{\perp i}^2 \approx
k_{\perp i}^2 \approx q_{\perp,i+1}^2$, is in eq.~(\ref{fLDC}) given a
weight
\begin{equation}
  \frac{\bar{\alpha}^2}{q_{\perp i}^2 q_{\perp,i+1}^2} \approx
  \frac{\bar{\alpha}^2}{\hat{t}^2}
\label{weight}
\end{equation}
where $\hat{t} \approx - k_{\perp i}^2$. This should be compared with
the well-known cross section for gluon-gluon scattering, which for
$\hat{s} >> |\hat{t}|$ is given by
\begin{equation}
  \frac{9}{2\pi} \frac{\alpha_s^2}{\hat{t}^2} = \frac{\pi^3}{2}
  \frac{\bar{\alpha}^2}{\hat{t}^2} 
\label{sigmaweight}
\end{equation}
The factor $1/k_{\perp 0}^2$ follows because $\mathcal{G}$ is defined
as a density in $\ln \kT^2$, rather than in $\kT^2$.

\subsection{Soft Chains}

It should be noted that $\sigma\sub{c}$ gives the cross section for a
chain with at least one link with virtuality larger than
$k_{\perp0}^2$.  There will also be chains with no perturbative
emissions above this cut.  (Emissions below the cut have to be added
later as final state radiation.)  The cross section for these soft
chains, i.e.\ the cross section for exchanging a single
(non-perturbative) gluon with virtuality $k_{\perp0}^2$ (see figure
\ref{fig:ppchain}b), is obtained by simply convoluting the two input
gluon densities:
\begin{equation}
  \label{eq:sigc0}
  \sigma\sub{c0}(k_{\perp0}^2) = \frac{\pi^3}{2k_{\perp0}^2}
  \int dx\sub{0+} dx\sub{0-}
  g(x\sub{0+},k_{\perp0}^2) g(x\sub{0-},k_{\perp0}^2)
  \delta(x\sub{0+}x\sub{0-}-k_{\perp0}^2/s).
\end{equation}
As we will see in section \ref{sect-results}, if the cutoff is
increased, the cross section for the perturbative chains will
decrease, but the cross section for the non-perturbative chains will
increase since the fitted input gluon density is then more divergent
for small $x$. This implies that the total number of chains in $pp$
collisions is insensitive to the cutoff $k_{\perp0}$, and therefore
can be fully determined by data on $F_2$, without any additional
adjustable free parameter.

The results in eqs.\ (\ref{sigmac}) and (\ref{eq:sigc0}) give together
with the total non-diffractive cross section the average number of
chains in one event. This average number is, however, not enough to
determine the properties of exclusive final states. The sub-collisions
may be correlated so that central collisions have a larger, and
peripheral a smaller, number of chains. These correlations will be
studied in more detail in section \ref{sect-MI}.

\FIGURE[t]{\scalebox{1.0}{\mbox{
\begin{picture}(340,280)(15,-20)
  \Line(40,20)(300,20)
  \Line(40,20)(170,280)
  \Line(170,280)(300,20)
  \Line(100,80)(77.5,35)
  \Text(100,90)[]{$q_{6}$}
  \Vertex(100,80){2}
  \Text(110,70)[]{$k_{5}$}
  \Vertex(125,80){2}
  \Text(125,90)[]{$q_{5}$}
  \Line(100,80)(125,80)
  \Line(125,80)(130,70)
  \Line(130,70)(145,70)
  \Text(140,60)[]{$k_{4}$}
  \Line(145,70)(160,100)
  \Line(160,100)(190,100)
  \Vertex(160,100){2}
  \Text(160,110)[]{$q_{4}$}
  \Text(175,90)[]{$k_{3}$}
  \Vertex(190,100){2}
  \Text(190,110)[]{$q_{3}$}
  \Line(190,100)(200,80)
  \Line(200,80)(230,80)
  \Text(215,70)[]{$k_{2}$}
  \Line(230,80)(247.5,45)
  \Vertex(230,80){2}
  \Text(230,90)[]{$q_{2}$}
  \Line(247.5,45)(260,45)
  \Text(250,40)[]{$k_{1}$}
  \Vertex(260,45){2}
  \Text(270,45)[]{$q_{1}$}
  \Line(260,45)(265,35)
  \LongArrow(250,180)(250,230)
  \LongArrow(250,180)(300,180)
  \Text(250,240)[]{$\ln \kTpot{2}$}
  \Text(310,180)[]{$y$}
  \DashLine(260,45)(260,20){2}
  \Line(260,45)(265,15)
  \Line(260,20)(265,15)
  \DashLine(100,80)(100,20){2}
  \Line(100,80)(110,10)
  \Line(100,20)(110,10)
  \DashLine(125,80)(125,20){2}
  \Line(125,80)(135,10)
  \Line(125,20)(135,10)
  \DashLine(160,100)(160,20){2}
  \Line(160,100)(180,0)
  \Line(180,0)(160,20)
  \DashLine(190,100)(190,20){2}
  \Line(190,100)(210,0)
  \Line(210,0)(190,20)
  \DashLine(230,80)(230,20){2}
  \Line(230,80)(240,10)
  \Line(240,10)(230,20)
  \Text(340,20)[]{$\ln p_{\perp\mbox{cut}}^{2}$}
  \DashLine(300,20)(310,20){2}
  \Text(330,55)[]{$\ln k_{\perp0}^{'2}$}
  \DashLine(35,55)(310,55){3}
  \Text(20,35)[]{$\ln k_{\perp0}^2$}
  \DashLine(35,35)(305,35){8}
  \DashLine(47.5,-20)(47.5,35){2}
  \DashLine(292.5,-20)(292.5,35){2}
  \LongArrow(200,-15)(292.5,-15)
  \LongArrow(140,-15)(47.5,-15)
  \Text(170,-15)[]{$\ln s/k_{\perp0}^2$}
  \DashLine(77.5,35)(77.5,-10){2}
  \Text(62.5,0)[]{\footnotesize $\ln\frac{1}{x_{0-}}$}
  \LongArrow(35,0)(47.5,0)
  \LongArrow(90,0)(77.5,0)
  \DashLine(265,35)(265,-10){2}
  \Text(280,0)[]{\footnotesize $\ln\frac{1}{x_{0+}}$}
  \LongArrow(305,0)(292.5,0)
  \LongArrow(252.5,0)(265,0)
  \LongArrow(220,130)(260,150)
  \Text(280,150)[]{$\ln q_+$}
  \LongArrow(120,130)(80,150)
  \Text(60,150)[]{$\ln q_-$}
\end{picture}}}
\vspace*{5mm}
  \caption{\label{fig:typ-chain} An example of a dipole chain in
    hadronic collisions. The initial-state emissions $q_i$ are marked
    with points in the $(y,\kappa=ln(q_{\perp}^2))$-plane. The
    connecting lines correspond to the propagators $k_i$.}}

\subsection{Final State Bremsstrahlung}

In the LDC model, final state radiation is emitted from the dipoles in
the chain in the same way as in the Dipole Cascade Model (DCM)
\cite{Gustafson:1986db,Gustafson:1988rq} for time-like parton
cascades, which is implemented in the \ariadne program
\cite{Lonnblad:1992tz}. In these time-like cascades the emissions are,
however, limited by the $k_\perp$ of the propagators in the initial
chain. In figure \ref{fig:typ-chain}, this corresponds to the area
below the line connecting the $q_i$ points.

The parameters in the DCM have been fitted, together with the
parameters in the string fragmentation \cite{Andersson:1983ia} in
\pythia, to reproduce a wide range of final-state observables at LEP
to an extraordinary precision. The main parameters in \ariadne are
$\Lambda\sub{QCD}$ and the cutoff in transverse momentum,
$p_{\perp\mbox{cut}}$, which have been fitted by the DELPHI
collaboration to $0.22$~GeV and $0.6$~GeV respectively
\cite{Hamacher:1995df}. The fact that $p_{\perp\mbox{cut}}$ typically
is smaller than the cutoff for the ISB, $k_{\perp0}$, obtained by
fitting LDC to \ftwo\ data, means that FSB will be allowed also below
$k_{\perp0}$ down to the hadronization cutoff, $p_{\perp\mbox{cut}}$.
This means that if $k_{\perp0}$ is increased to $k'_{\perp0}$, some
gluon emissions in the initial state, such as $q_1$ in figure
\ref{fig:typ-chain}, will no longer be allowed, but will then instead
be allowed as FSB. In this way the distribution of the final state
partons should be quite insensitive to the precise value of
$k_{\perp0}$.

Special care must be taken when radiating from the dipoles connecting
the first ISB gluons with the hadron remnants. This problem is
especially important for the soft chains, where FSB will also be
allowed below $k_{\perp0}$ from a dipole between the remnants. This
will be discussed further in section \ref{sec-mc}.

\section{Multiple scatterings in hadronic collisions}
\label{sect-MI}

At high energies the inclusive cross section for parton-parton
sub-collisions becomes larger than the total non-diffractive cross
section, which means that there are on average more than one hard
collision per event. Indications for multiple hard collisions are
observed both in hadronic collisions \cite{Sjostrand:1987su} and in
photoproduction \cite{Butterworth:1996zw}. The properties of exclusive
final states then depend sensitively on the distribution in the number
of sub-collisions in a single event. At high energies two colliding
hadrons are Lorentz contracted to two flat pancakes. When they pass
through each other the different regions are causally disconnected,
and it is then natural to assume that the sub-collisions are
independent for a fixed value of the impact parameter, $b$. This
implies an eikonalized description, where the probability to have $n$
sub-collisions is given by a Poissonian distribution with an average,
$\bar{n}(b)$, which depends on $b$, or alternatively by such a
distribution with the value $n=0$ excluded. Models of this type
include \pythia, \herwig \footnote{Using the add-on \jimmy program
  \cite{Butterworth:1996zw}}, and the Dual Parton Model.

\subsection{Impact-Parameter Dependence}

We note that the result depends not only upon the average number of
hard sub-collisions for fixed impact parameter, $\tilde{n}(b)$, but
also on the probability for an interaction, $P_{int}(b)$, for this
value of $b$. These quantities are defined so that the inclusive
sub-collision cross section is given by $\sigma_{hard} = \int d^{2}b\,
\tilde{n}(b)$ and the total non-diffractive cross section by the
relation $\sigma_{tot} = \int d^{2}b\, P_{int}(b)$. Expressed in these
quantities we have
\begin{equation}
\bar{n}(b) = \tilde{n}(b) / P_{int}(b)
\label{eq:A} 
\end{equation}

In ref.~\cite{Sjostrand:1987su} it is found that e.g.\ the "pedestal
effect", i.e.\ the fact that the background is larger around a high
\ET\ jet than in a minimum bias event, and the distribution in the
number of jets in a single event can be well reproduced if the
$b$-dependence is described by a double Gaussian distribution,
corresponding to the presence of a kind of "hard core", while a single
Gaussian would severely underestimate the fluctuations. (By a Gaussian
distribution we understand here a distribution which is Gaussian in
the variable $b$, and consequently exponential in the density $b^2$.)

In this paper we will not present a full investigation of such
correlations between sub-collisions, but only study some qualitative
features.  The fact that there often are more than two jets in one
single chain in our approach, will imply a positive correlation
between the sub-collisions in addition to the one originating from the
impact parameter dependence.  Therefore the fluctuations in the number
of chains due to the impact parameter dependence should be weaker than
the fluctuations in the number of sub-collisions in the models
mentioned above. To illustrate the effect of these two contributions
to the correlations, we will study two different cases:
\begin{itemize}
\item Completely independent chains, where the number of chains is
  determined by a Poissonian distribution
  \begin{equation}
    \label{eq:pois}
    P_n = e^{-\bar{n}} \frac{\bar{n}^n}{n!}
  \end{equation}
  where $\bar{n}$ is the $b$-independent average number of chains in
  an event.
\item An impact parameter dependence which results in an exponential
  distribution in $\bar{n}(b)$ when weighted by the probability
  $P_{int}(b)$. The probability $P(n')$ for a collision with a
  centrality corresponding to $\bar{n}(b)=n'$ is thus assumed to
  satisfy
  \begin{equation}
    P(n') = \frac{1}{\sigma_{tot}} \int d^{2}b P_{int}(b) \delta (\bar{n}(b)
    - n') = \bar{n} e^{-n'/\bar{n}}
  \end{equation}
  Here $\bar{n}$ is the average $n'$-value. Assuming a Poissonian
  distribution in $n$ for fixed value of $b$, i.e.\ for fixed value of
  $n'$, we then find a geometric distribution with the same average,
  $\bar{n}$, for the number of chains, $n$, in a random event.  For
  the distribution in the number of chains we then find the geometric
  distribution in $n$, with the same average, $\bar{n}$:
  \begin{equation}
    \label{eq:geom}
    P_n = \int d n' P(n') \cdot e^{-n'} \frac{n'^n}{n!} =
    \frac{1}{1+\bar{n}} (\frac{\bar{n}}{1+\bar{n}})^n
  \end{equation}
\end{itemize}
\FIGURE[t]{ \epsfig{figure=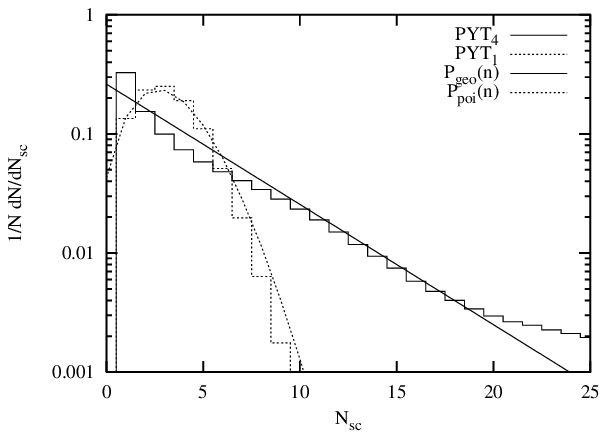,width=10cm}

  \caption[dummy]{\label{fig:geom} The distribution in number of
    scatterings for non-diffractive events in \pythia. The full
    histogram is \PYTD and the dotted histogram is \PYTP (see
    subsection \ref{subsect-fit} for notation). The full and dotted
    curves are geometrical and Poissonian distributions respectively
    with the same average as the corresponding histograms.}}

In figure \ref{fig:geom} we compare the simple geometrical
distribution with the distribution in number of scatterings in \pythia
with a double Gaussian $b$-distribution, and we find that the
geometrical distribution actually is extremely similar. Also shown is
\pythia with independent scatterings and a Poissonian distribution for
comparison.

A reason to concentrate on these two alternatives in this preliminary
study, is that the Poissonian distribution will act as a reference for
the effect of further correlations, while the geometric distribution
can mimic the fluctuations in the favoured \pythia version.  We also
note that both these distributions are chosen so that they depend only
on a single parameter, the average number of chains, $\bar{n}$.

\section{Monte Carlo implementation}
\label{sec-mc}

The LDC event generator \ldcmc, is written to model deeply inelastic
lepton--hadron scattering or, more specifically, $\gamma^*$-$h$
scattering. To use it for generating single dipole chains in hadronic
collisions we lower the virtuality of the $\gamma^*$ to below
$k_{\perp0}$ and use a reweighting procedure to obtain the correct
input parton densities for the hadron and to account for the fact that
the photon only couples to quarks, while the hadron also has a gluon
density.

Since we expect that there will be several chains in each hadronic
collision, we need to discuss two problems concerning how to combine
the generated chains: we need to consider the colour connections
between the chains and the hadron remnants, and we need to ensure
energy--momentum conservation, so that the chains do not take more
than the available energy in the colliding hadrons.

\subsection{Colour Connections}

We do not expect all individual chains to be connected to the hadron
remnant -- this would give a large set of overlapping strings
producing too many hadrons in the full rapidity region. Instead we
expect that ISB partons in the ends of each chain may be connected to
corresponding partons in other chains. One could also imagine
situations where two chains are different branches on the same
cascading tree. This possibility will, however, be neglected in the
present analysis. To be less sensitive to the colour connections we
will here only consider the properties of the \emph{partonic} final
state. However, although less sensitive, also the final state
radiation will be affected to some extent, since the emitting dipoles
depend on these connections.

Clearly, the question of colour connections is a very complicated one,
and we will defer a detailed investigation to a later publication.
Here we will simply assume that the FSB can be added to each chain
independently, also allowing the dipoles between the ISB partons in
the ends of the chain and the hadron remnants to radiate. This will
certainly somewhat overestimate the amount of soft partons in the
events and the results presented below must therefore be considered
somewhat preliminary.

The FSB emissions from the remnant dipoles also pose another problem.
The emission probability from a dipole is basically flat in $\ln
p_\perp^2$ and rapidity. This means that the emitted gluon can easily
take a large fraction of the remnant momentum. The standard way of
dealing with this in \ariadne is the Soft Radiation Model
\cite{Andersson:1989gp}, but here we will use a different method, to
be as insensitive as possible to the cutoff, $k_{\perp0}$. Looking at
the initial-state emission of the gluon $q_1$ in figure
\ref{fig:typ-chain}, it will also have a basically flat rapidity
distribution, except there will be a $(1-x_{0+})^4$ suppression from
the input gluon density (see eq.~(\ref{eq:fit-form}) and table
\ref{tab:param} in section \ref{subsect-fit}). If we now increase the
cutoff to $k'_{\perp0}$, $q_1$ should instead be allowed as a final
state emission with approximately the same distribution. We have
therefore chosen a procedure where for each gluon emission generated
according to the Dipole Cascade Model in a remnant dipole, a
corresponding new $x'_{0+}$ is calculated from the energy loss of the
remnant, and the emission is only accepted with a probability
\begin{equation}
  \label{eq:remdipveto}
  \left(\frac{1-x'_{0+}}{1-x_{0+}}\right)^4.
\end{equation}
The correct procedure would, of course, be to use the actual ratio of
the relevant input parton densities. We have, however, checked that
the results are insensitive to the details of the suppression (e.g.\ 
changing the exponent from 4 to 3) and will therefore in this paper be
satisfied with this simplified treatment.

\subsection{Energy--Momentum Conservation}

The treatment of emissions from the remnant dipole is also related to
the question of energy--momentum conservation when combining several
chains. If each chain takes a large fraction of the energy of the
incoming hadron, there will not be enough energy to have many chains.
To ensure energy--momentum conservation we have adopted the following
procedure when combining chains into single events.
\begin{enumerate}
\item First we choose $n$, the number of chains to be included in the
  event, from either the Poissonian distribution in
  eq.~(\ref{eq:pois}) or the geometrical one in eq.~(\ref{eq:geom}).
\item Then we choose $n$ chains, either hard chains with both ISB and
  FSB partons or soft ones with only FSB, according to the values of
  $\sigma_c$ and $\sigma_{c0}$.
\item The selected chains are then ordered in the transverse momentum
  of the hardest parton.  This procedure is analogous to the treatment
  in the \pythia model. It implies that the hardest chain is always
  accepted in the following veto procedure, which is necessary to
  ensure that the spectrum of high \ET\ jets is correctly described.
\item Each chain is now tested in order of decreasing hardness. The
  sum of the positive and negative light-cone momentum already used up
  by the previously accepted chains is calculated and from that, new
  effective values of $x'_{0\pm}$ of the current chain are obtained.
  Chains which would take more than the available energy are vetoed.
  Kinematically allowed chains are accepted with a probability
  \begin{equation}
    \label{eq:energyveto}
    \left(\frac{1-x'_{0+}}{1-x_{0+}}\times
      \frac{1-x'_{0-}}{1-x_{0-}}\right)^4,
  \end{equation}
  to approximate the change in the input parton densities after some
  partons have already been extracted. As for the treatment of remnant
  dipoles in eq.~(\ref{eq:remdipveto}) it may be more correct to here
  use the ratio of the actual input densities used. (Again we have
  checked that our results are not very sensitive if the exponent is
  changed from 4 to~3.)
\item We are now left with $n\sub{acc}\leq n$ accepted chains which
  are combined into a partonic event.
\end{enumerate}
 
The procedure presented here is at present rather cumbersome and slow
and some results presented here have rather large statistical
fluctuations. The absence of hadronization also makes the current
version of the program unsuited for direct comparison with experiment.
A new publicly available version of \ldcmc capable of generating
hadronic collision events will therefore be postponed until the issues
of colour connections have been addressed in more detail.

\section{Results}
\label{sect-results}

In this section we will study features of jet and minijet production,
which are sensitive to the properties of multiple interactions.  We
will look at inclusive jet production, minijet distributions, and the
pedestal effect and other properties of underlying events.

As example we have chosen $pp$ collisions at the Tevatron energy,
1.8~TeV in the cms. To be specific we define jets with a simple
cone-algorithm\footnote{\pycell in \pythia} using an idealized
calorimeter with cells distributed uniformly in pseudo-rapidity and
azimuth angle with the size
$\delta\eta\times\delta\phi\approx0.2\times0.2$. The transverse energy
of all generated partons are collected in the corresponding cells, and
any cell with a summed \ET\ above 1~GeV is tried as a jet-initiator.
If the summed \ET\ within a radius of 0.7 in the $\eta$-$\phi$ plane
around such an initiator is above 3~GeV, this \ET\ is assigned to a
jet in the \ET\ weighted direction of the cells included. Only jets
with $|\eta|<2.5$ are included in the following.

As discussed above we will in this preliminary study not include
hadronization effects, which are sensitive to how the chains are
colour connected. We will therefore only compare with parton
distributions generated by the \pythia generator, and make no direct
comparisons with experimental data.

\subsection{Fitting procedure}
\label{subsect-fit}

\TABLE[t]{
  \begin{tabular}{|l||r|r|r||r|r||r|r||r|r||r||r|}
    \hline
    fit & $A_g$ & $a_g$ & $b_g$ &
    $a_d$ & $b_d$ & $a_u$ & $b_u$ & $a_s$ & $b_s$ & $k_{\perp 0}$ &
    $\chi^2/\mbox{d.o.f.}$ \\
    \hline
    fit-0 & 2.24 & 0.09 & \textbf{4} &
    1.77 & \textbf{3} & 0.58 & \textbf{3} &
    \textbf{0} & \textbf{4} & \textbf{0.88} & 795/625 \\
    fit-1 & 1.86 & 0.00 & \textbf{4} &
    1.78 & \textbf{3} & 0.57 & \textbf{3} &
    \textbf{0} & \textbf{4} & 0.99 & 694/625 \\
    fit-2 & 1.42 & -0.09 & \textbf{4} &
    1.94 & \textbf{3} & 0.56 & \textbf{3} &
    \textbf{0} & \textbf{4} & \textbf{1.14} & 710/625 \\
    fit-3 & 1.17 & -0.14 & \textbf{4} &
    2.10 & \textbf{3} & 0.55 & \textbf{3} &
    \textbf{0} & \textbf{4} & \textbf{1.3} & 751/625 \\
    \hline
  \end{tabular}
  \caption{The result of the fit of the parameters for the
    input parton densities to data from H1 \cite{Aid:1996au}, ZEUS
    \cite{Derrick:1996hn}, NMC \cite{Arneodo:1995cq} and E665
    \cite{Adams:1996gu} in the region $x<0.3$, $Q^2>1.5$~GeV$^2$.  The
    last columns gives the $\chi^2$ over the number of fitted data
    points, respectively. Parameters in bold face have not been
    fitted.}
  \label{tab:param}}

The parameters used in the MC simulations are fitted to DIS data.  To
get quantitative results from LDC we must convolute the dipole chains
with input parton densities in the proton. These densities are not
\`{a} priori known, but must be parameterized in some way and fitted
to data. The form of these densities are taken to be
\begin{equation}
  xf_i(x,k_{\perp0}^2) = A_i x^{a_i} (1-x)^{b_i},
  \label{eq:fit-form}
\end{equation}
where $i=d_v,u_v,g$ and $s$ for the d-valence, u-valence, gluon and
sea-quark densities respectively (where the sea flavour densities are
assumed to be $f_{\bar{d}}=f_{\bar{u}}=2f_{\bar{s}}$). The parameters
$A_i,a_i,b_i$ and the perturbative cutoff, $k_{\perp0}$, are then
fitted to reproduce the measured data on $F_2$. There are some
relations between the parameters which are fixed by sum rules. Thus
$A_{d_v}$ and $A_{u_v}$ are determined by flavour conservation and
$A_s$ is fixed by momentum conservation. The fits to $F_2$ do not
constrain the remaining parameters very strongly, so we have fixed the
$b$ parameters to $3$ in the valence densities and to $4$ in the sea
and gluon densities. The best fit is the one called \textit{standard}
in reference \cite{Gustafson:2002jy} and fit-1 in table
\ref{tab:param}.  The fitted value of $k_{\perp0}$ is 0.99~GeV, but
since we would like to check the dependence on the cutoff, we also
present additional fits in table \ref{tab:param} where $k_{\perp0}$
has been fixed to higher and lower values.

In the following we will look at a couple of options for LDC and
compare with different cases in \pythia.  (For the alternatives with a
single chain (\LDCz and \LDCzp) or a single sub-collision (\PYTz) we
can only study inclusive cross sections, and the reduction from energy
conservation due to overlayed chains or collisions is not included.)

\begin{description}
  \itemsep 0mm
\item[\LDCz:] Single LDC chains with the default (fit-1) tuning.
\item[\LDCP:] Uncorrelated multiple LDC chains (fit-1) according to
  the Poissonian distribution.
\item[\LDCG:] Multiple LDC chains (fit-1) according to the geometrical
  distribution.
\item[\LDCzp:] As \LDCz but with the $k_{\perp0}=1.3$~GeV tuning
  (fit-3) (similarly for \LDCGp).
\item[\PYTz:] \pythia without multiple scatterings selecting low-\pT\ 
  events without hadronization. Everything else default.
\item[\PYTP:] As \PYTz but with uncorrelated multiple scatterings
  (with default values\footnote{The indices for the \pythia options
    refer to the setting of the switch, \texttt{MSTP(82)}, selecting
    the multiple interaction model.}).
\item[\PYTD:] As \PYTz but with multiple scatterings according to a
  double Gaussian distribution (with default values).
\item[\PYTDp:] As \PYTD but with a higher soft regularization
  parameter (2.5~GeV instead of 2.1~GeV).
\end{description}

Since \PYTD is known to give a fair description of experimental data,
we will in the following use it as a reasonable approximation to
nature. We should keep in mind, however, that the results presented
here are on parton level only, and that many of the observables
studied have not been measured experimentally, so deviations from
\PYTD do not imply that our model is wrong. We will also mostly
compare \LDCP with \PYTP, which both are based on the assumption of
uncorrelated scatterings; and \LDCG with \PYTD, since we have shown in
figure \ref{fig:geom} that \PYTD is well approximated by a geometrical
distribution. Note, however, that the procedure for energy
conservation in LDC severely limits the possibility to have many
chains, and \LDCG deviates substantially from a geometrical
distribution, as can be seen in figure \ref{fig:geomldc}. The effects
of energy conservation in \pythia is much smaller since the ISB is
modeled with a DGLAP parton shower with strongly ordered virtualities
which means that only very rarely are partons emitted close to the
hadron remnants.

\FIGURE[t]{ \epsfig{figure=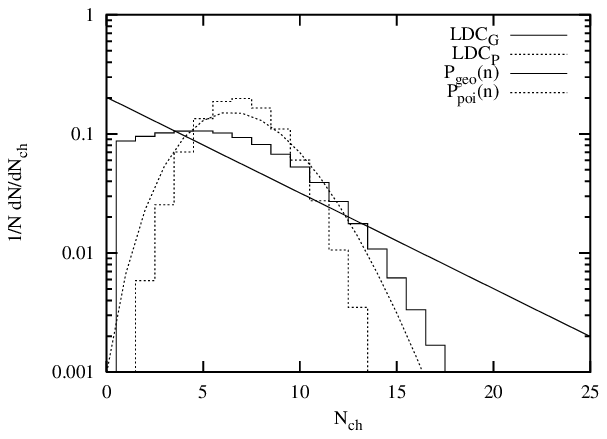,width=10cm}

  \caption[dummy]{\label{fig:geomldc} The distribution in number of
    chains per event in LDC. The full histogram is \LDCG and the
    dotted histogram is \LDCP. The full and dotted curves are
    geometrical and Poissonian distributions respectively with the
    same average as the corresponding histograms.}}

\subsection{Chain multiplicity}
\label{subsect-chain}

\FIGURE[t]{ \epsfig{figure=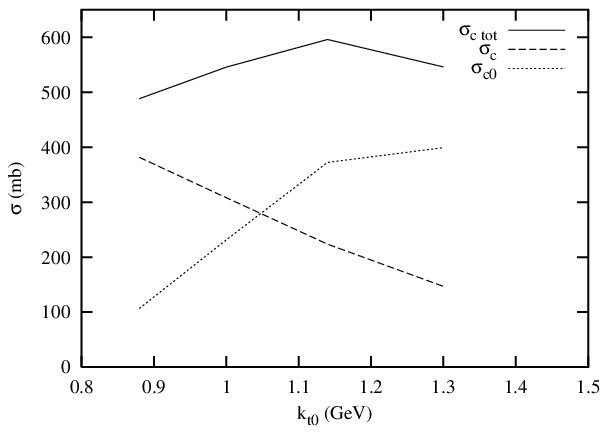,width=10cm}
  \caption{\label{fig:sigkt0} The cross section per chain in the LDC
    model as a function of the cutoff, $k_{\perp0}^2$. The dashed line
    is the cross section for chains with at least on emission above
    the cutoff, the dotted line is for soft chains without emissions
    above the cutoff, and the full line is the sum of the two. Note
    that the input parton densities have been re-fitted for each value
    of $k_{\perp0}$.}}

First we want to study the cross section for the formation of a chain
in a hadronic collision, which is given by eqs.~(\ref{sigmac}) and
(\ref{eq:sigc0}).  For fit-1 at a c.m.\ energy of 1.8~TeV, we obtain
for perturbative chains a cross section of $\sigma\sub{c}=315$~mb,
which is much larger than the non-diffractive cross section at the
Tevatron\footnote{We take the value as estimated by a simple
  parameterization in \pythia based on ref.~\cite{Donnachie:1992ny}.},
$\sigma\sub{nd}\approx39$~mb. In figure \ref{fig:sigkt0} we see how
this cross section per chain varies with the soft cutoff used in the
fit to DIS $F_2$ data. The result is an almost linear dependence on
the cutoff $k_{\perp0}$. (Note that the input parton densities have
been re-fitted for each value of $k_{\perp0}$.)  In figure
\ref{fig:sigkt0} we also show the cross section for formation of soft
chains, represented by the exchange of a single (non-perturbative)
gluon with virtuality $k_{\perp0}^2$ (cf.\ eq.~(\ref{eq:sigc0})).  We
see that if the cutoff is increased, the cross section for the
perturbative chains will decrease, but the cross section for the
non-perturbative chains will increase since the fitted input gluon
density is then more divergent for small $x$. From figure
\ref{fig:sigkt0} we see that the sum of the two cross sections is
almost independent of the cutoff $k_{\perp0}$.  Thus we conclude that
the total number of chains in $pp$ collisions can be fully determined
by data on $F_2$, without any additional adjustable free parameter.

Taking the sum of $\sigma_c$ and $\sigma_{c0}$ (547~mb for
$k_{\perp0}=0.99$~GeV) we would then expect on the average around 14
chains per collision. However, after the veto procedure to conserve
energy and momentum we are left with an average of 7 and 6 chains per
event for \LDCP and \LDCG respectively.

\subsection{Inclusive jet production}
\label{subsect-inclus}

\FIGURE[t]{ \epsfig{figure=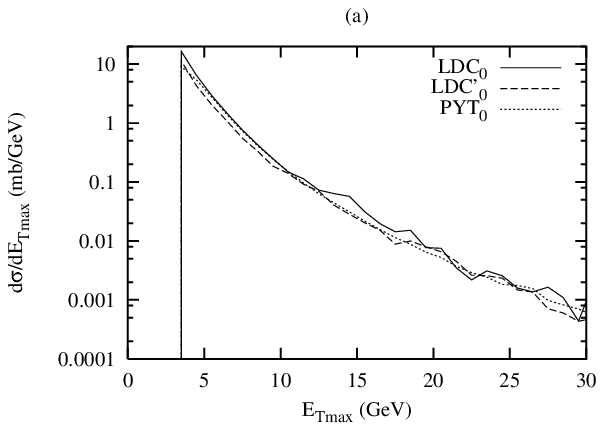,width=7.4cm}
  \epsfig{figure=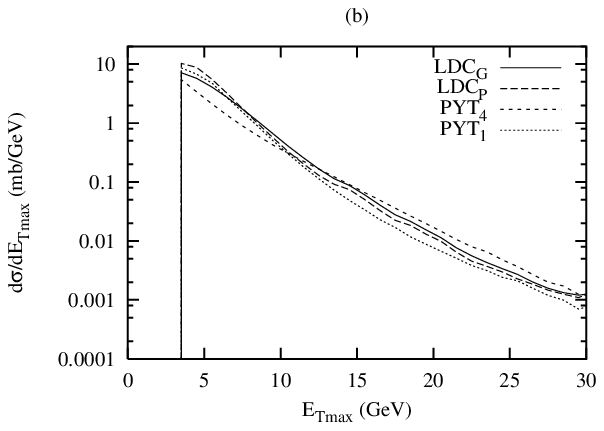,width=7.4cm}
  \caption{\label{fig:etmax0} The differential \ETmax\ cross
    section. In (a) \ETmax\ is the largest \ET\ jet in a chain or hard
    scattering. Full line is \LDCz, dashed is \LDCzp and the dotted
    line is \PYTz. In (b) \ETmax\ is the largest \ET\ in an event.
    Full line is \LDCG, long-dashed is \LDCP, short-dashed is \PYTD
    and the dotted line is \PYTP}}

In figure \ref{fig:etmax0}a we look at the inclusive differential
cross section for the highest \ET\ jet in a single LDC chain and
compare it to the corresponding cross section in \pythia without
multiple interactions (\PYTz).  We find that the \ldcmc result is
close to \pythia. Comparing \LDCz and \LDCzp we find a weak dependence
on the cutoff. The reason for this may be that the input densities in
LDC has been fitted to \ftwo\ data which are only indirectly sensitive
to the gluon contribution. Indeed, in \LDCzp the gluons carry a
smaller fraction of the proton momentum which could account for the
differences in figure \ref{fig:etmax0}a. In the future, when \ldcmc
can be compared directly with high-\ET\ jet production, it should be
possible to make a more global fit of the input parton densities,
which would better constrain the gluon distribution. Another reason
for the dependence could be the differences in FSB caused by our
simplified treatment of the colour connections of the chains.

In figure \ref{fig:etmax0}b we show the differential \ETmax\ cross
with multiple interactions added with Poissonian (\LDCP) and
geometrical (\LDCG) distributions for the number of chains in one
event. For comparison we also show \PYTP and \PYTD. We find that the
LDC curves more or less lie in between the \pythia ones, which
indicates that the procedure chosen to combine chains is a reasonable
one. We note, however, a difference for small \ETmax\ values where
\LDCG lies significantly above \PYTD.

\FIGURE[t]{ \epsfig{figure=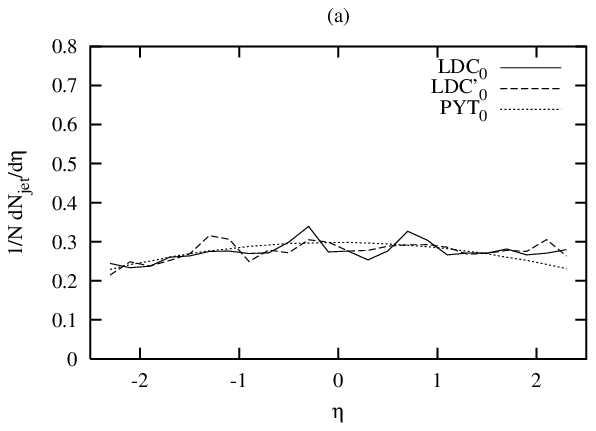,width=7.4cm}
  \epsfig{figure=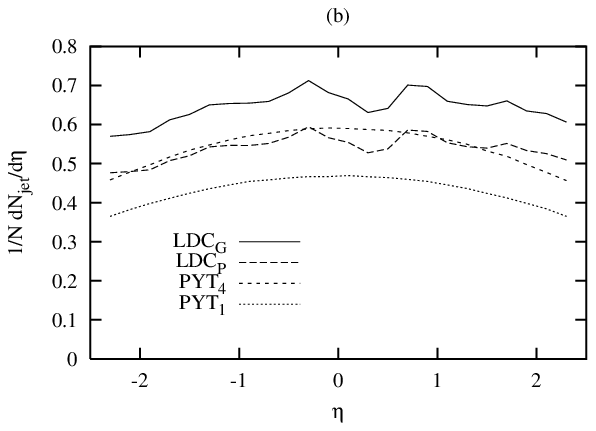,width=7.4cm}
  \caption{\label{fig:rap} The rapidity distribution of minijets 
    (per event with at least one minijet) In (a) the full line is
    \LDCz, dashed is \LDCzp and the dotted line is \PYTz. In (b) the
    full line is \LDCG, long-dashed is \LDCP, short-dashed is \PYTD,
    dash-dotted is \PYTS and the dotted line is \PYTD. (The different
    LDC curves have been obtained using the same sample of chains
    which is why the statistical fluctuations are correlated.) } }

\subsection{Minijet distributions}
\label{subsect-minijet}

As another check that our procedure is reasonable, we look at the
rapidity distribution of minijets. In figure \ref{fig:rap}a we show
the results for \LDCz, \LDCzp and \PYTz. We find that the curves agree
well, and the cutoff dependence in LDC is negligible. The slight
asymmetry in the LDC distributions is due to the fact that the recoil
strategy for the final state emissions from the parton closest to the
photon in the \ldcmc for DIS is not fully consistently modified in
this preliminary application to hadronic collisions.

In figure \ref{fig:rap}b we show the results for the multiple
interaction models. Here the LDC curves are above the corresponding
\pythia ones. Although there are no obvious reasons for the curves to
agree, we note that the soft FSB in LDC may be somewhat overestimated,
especially for the soft chains. We also note that the statistics for
the LDC curves is poor due to the present inefficient simulation
procedure.  In addition, the statistical fluctuations in \LDCP and
\LDCG are correlated since the two curves are obtained using the same
set of generated single LDC chains.

\FIGURE[t]{ \epsfig{figure=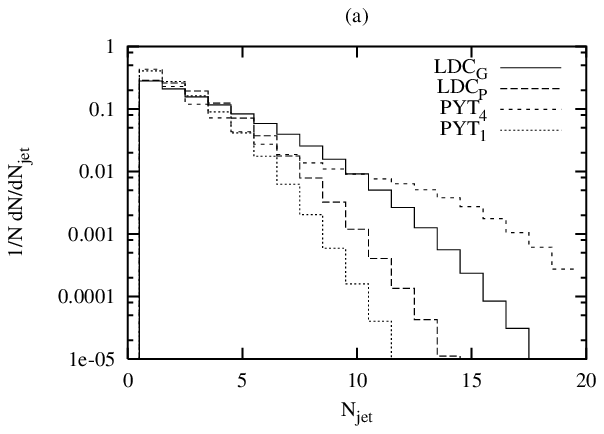,width=7.4cm}
  \epsfig{figure=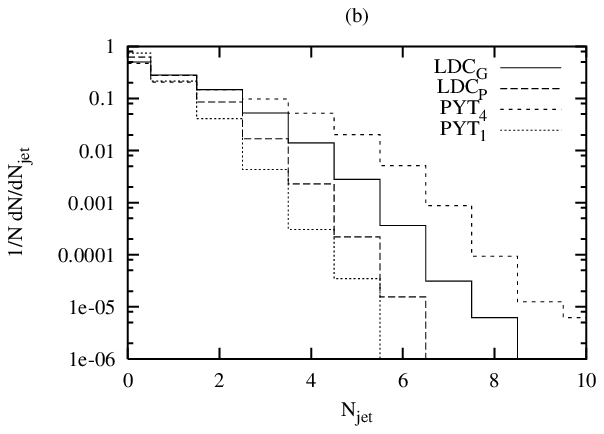,width=7.4cm}
  \caption{\label{fig:njet} The distribution in number of minijets 
    in the range $|\eta|<2.5$ (per event with at least one minijet).
    (a) shows all minijets, while (b) is all minijets in the minimum
    azimuth region (as defined in the text) in each event with
    $\ETmax>10$~GeV. In both cases the full line is \LDCG, long-dashed
    is \LDCP, short-dashed is \PYTD, and the dotted line is \PYTP.}}

To investigate the fluctuations we show in figure \ref{fig:njet}a the
distribution in number of jets in the central 5 units of rapidity for
the different multiple interaction models. It is clear that \LDCG has
much larger fluctuations than has \LDCP, as was expected. \LDCP has
also larger fluctuations than \PYTP although both are based on
uncorrelated scatterings. This is because in LDC there is, as
mentioned above, a possibility to have several hard scatterings in
each chain, which increases the correlations. \LDCG does not have as
large fluctuations as \PYTD, which also was expected since, as shown
in figure \ref{fig:geomldc}, the distribution in number of chains in
\LDCG is much narrower than a geometrical distribution, due to energy
conservation.

\subsection{The Pedestal Effect and Underlying events}
\label{subsect-underlying}

The contribution from hard sub-collisions is in many analyses added on
top of an independent ``underlying event''.  In reference
\cite{Affolder:2002zg} a number of observables were presented which
were shown to be especially sensitive to this background activity. The
regions of azimuth which are transverse to the highest \ET\ jet in an
event given by $60^\circ<\phi<120^\circ$, should be less affected by
the primary \emph{trigger} scatterings, which mainly populate the
collinear and opposite azimuth region from the highest \ET\ jet with
hadrons. There are, however, also effects from bremsstrahlung from the
primary interaction in the transverse regions, but the authors argue
that these typically only affect one of these two regions.  Thus, by
looking at the activity in the transverse region which has the
smallest activity, one is especially sensitive to effects of the
``underlying event'' and secondary scatterings. We will call this
region the \textit{minimum azimuth region}, and in \ref{fig:njet}b we
show the distribution in number of jets in this region in events with
a trigger jet of $\ETmax>10$~GeV. Clearly, the general trend found for
the overall distribution in number of jets is present also here.

\FIGURE[t]{ \epsfig{figure=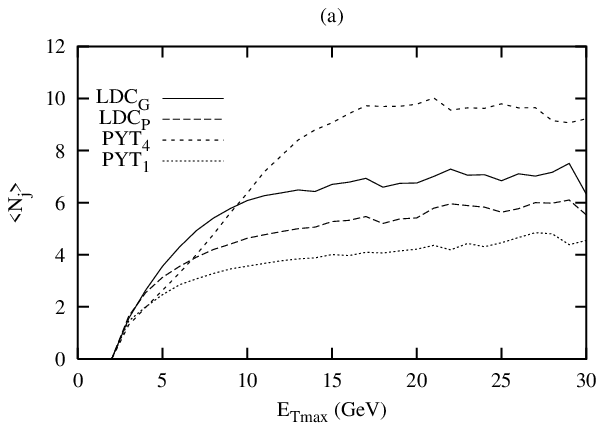,width=7.4cm}
  \epsfig{figure=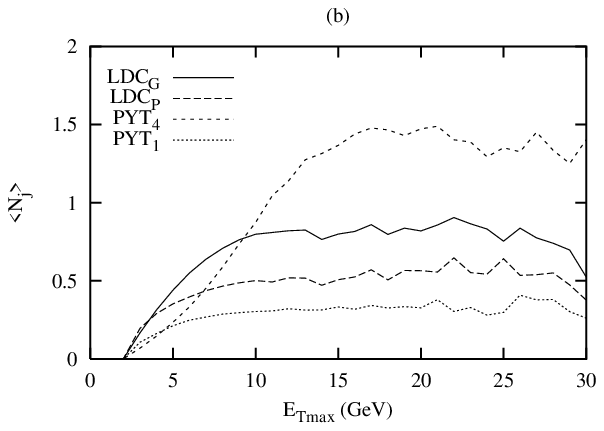,width=7.4cm}
  \caption{\label{fig:avjet} (a) The average number number of minijets
    per event as a function of the \ET\ of the largest jet in the
    event. The full line is \LDCG, long-dashed is \LDCP, short-dashed
    is \PYTD and the dotted line is \PYTP. (b) Is the same but looking
    only at minijets in the minimum azimuth region.}}

One of the most important features of the underlying event is the
so-called jet pedestal effect, i.e.\ the fact that the underlying
event activity is not independent of a large \ET\ jet. In figure
\ref{fig:avjet} we show the average number of minijets as a function
of the \ET\ of the hardest jet, both for the overall minijet activity
and for the minimum azimuth region. As expected all models show an
increasing average number with increasing \ETmax\ since, even in the
case of uncorrelated scatterings, there is a higher probability to
find a large \ET\ jet in an event with many scatterings than in an
event with few. Also, the models with most fluctuations gives the
strongest rise, while models with fewer fluctuations saturate earlier.
We note, however, that for this variable there is a particularly large
difference between the \LDCG and \PYTD models.

\subsection{Cutoff Dependence}

\FIGURE[t]{ \epsfig{figure=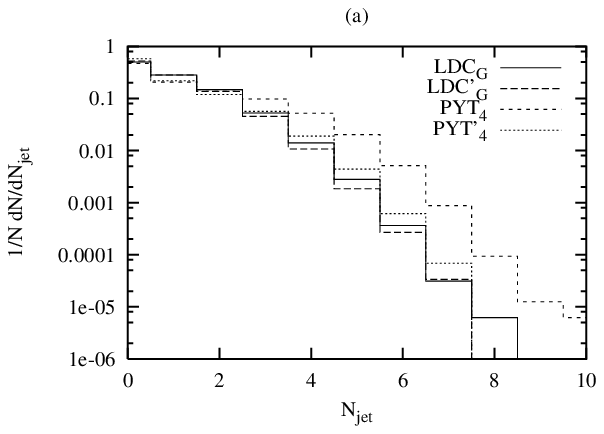,width=7.4cm}
  \epsfig{figure=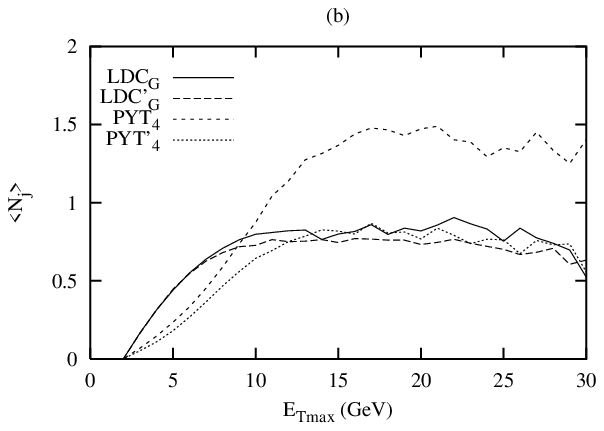,width=7.4cm}
  \caption{\label{fig:cutdep} (a) The distribution in number of
    minijets in the minimum azimuth region in events with
    $\ETmax>10$~GeV. (b) The average number of minijets per event in
    the minimum azimuth region as a function of the \ETmax.  In both
    plots the full line is \LDCG, long-dashed is \LDCGp, short-dashed
    is \PYTD and the dotted line is \PYTDp.}}

Finally we would like to check the cutoff dependence also for our
multiple scattering model. In figure \ref{fig:cutdep} we show the
distribution in number of jets and the average number of jets as a
function of \ETmax\ for \LDCG and \LDCGp and we find that the
differences indeed are very small. As comparison we also show the
results for \PYTD and \PYTDp. For the double-Gaussian model in
\pythia, there is no direct cutoff, but rather a soft regularization
parameter, $p_{\perp0}$, which forces the partonic cross sections and
the running of \as\ to be finite at small scales by modifying them
according to
\begin{equation}
  \label{eq:pytcut}
  \frac{d\sigma\sub{part}(p^2_{\perp})}{dp^2_{\perp}}\rightarrow
  \frac{d\sigma\sub{part}(p^2_{\perp})}{dp^2_{\perp}}\times
  \left(\frac{p^2_{\perp}}{p^2_{\perp}+p^2_{\perp0}}\right)^2,\,\,\,\,\,\,
  \as(p^2_{\perp})\rightarrow\as(p^2_{\perp}+p^2_{\perp0})
\end{equation}
We see in figure \ref{fig:cutdep} that the result is very sensitive to
the value of $p_{\perp0}$, which has to be adjusted to fit
experimental data.  We also note that in \pythia the value of
$p_{\perp0}$ is expected to increase with energy in a way which is not
determined by the model, which means that it is difficult to make
reliable predictions for future experiments at higher energies.
Naturally we should not compare the variation in cutoff for LDC
($0.99\rightarrow1.3$~GeV) with the variation in the regularization
parameter in \pythia ($2.1\rightarrow2.5$~GeV).  From figure
\ref{fig:cutdep} we can, however, see that the dependence in LDC is
very weak, which means that in this model the result can be directly
estimated from data from DIS.

\section{Conclusions}
\label{sect-conclusions}

In this article we present a new approach to describe high energy
hadronic collisions, which is based on the Linked Dipole Chain model
for DIS.  The main advantage over earlier treatments is a much reduced
sensitivity to an infrared cutoff, which in models based on collinear
factorization has to be fitted to experimental data. In our model the
parton-parton sub-collision cross section can be fully determined by
data from DIS, which implies a more intimate relation between
hadron-hadron collisions and DIS. We hope that this relation also
implies increased possibilities to use data on $hh$ collisions to
constrain fits to parton distribution functions.  Thus e.g.\ the
connections between different parton chains in $hh$ collisions may be
related to gluon saturation effects in DIS. The insensitivity to a
soft cutoff also makes it easier to make reliable extrapolations to
higher energies.

At high energies the properties of hadron-hadron collisions appear to
be dominated by hard and semi-hard parton collisions. Although the
parton-parton collisions are of perturbative nature, various
non-perturbative effects make a description of hadronic collisions
much more complicated than e.g.\ DIS:

\begin{itemize}
\item[(i)] In a traditional collinear factorization scheme the
  inclusive sub-collision cross section is divergent for small $\pT$,
  which has to be treated by a soft cutoff $p_{\perp 0}$.
  
\item[(ii)] The inclusive sub-collision cross section is larger than
  the total non-diffractive cross section, which means that there may
  be many hard sub-collisions in a single event. This implies that the
  correlations between the sub-collisions are important.
  
\item[(iii)] The scattered partons can be colour connected with each
  other and with the proton remnants in different ways by strings or
  cluster chains.
\end{itemize}

In models based on collinear factorization the soft cutoff, $p_{\perp
  0}$, has to be fitted to experimental data. The correlation between
partonic sub-collisions is, in e.g.\ the DPM model, \pythia, and \herwig,
described in an eikonalized form by an impact parameter dependence.
For a single hard sub-collision the colour structure can be read off
from the perturbative matrix elements
\cite{Gustafson:1982ws,Bengtsson:1984jr}, but for events with many
sub-collisions the result necessarily depends on non-perturbative
effects.

The LDC model for DIS is a reformulation and generalization of the
CCFM model. It is based on $\kT$-factorization and the inclusion of
non-$\kT$-ordered parton chains, which are expected to be important
for limited $\pT$ and high energies. The formulation is fully
symmetric with respect to the photon and the proton ends of the parton
chain, which makes the model also suitable for a description of
hadronic collisions.  The input soft gluon distribution,
$g_0(x,Q_{0}^2)$, is fitted to data from DIS, and when we then
calculate the number of parton chains in $pp$ collisions, we find that
the result is actually fixed by the fit to $F_2$. If the soft cutoff
used in the fit is increased the number of hard chains is reduced, but
this is fully compensated by an increase of soft chains described by
the input distribution function, $g_0$. (These soft chains are now
also allowed to emit final state radiation up to the increased value
for the $k_\perp$-cutoff.)

Concerning point (ii) in the list above, we note that some
correlations between hard sub-collisions are included automatically,
as a single non-$\kT$-ordered parton chain can contain two or more
local maxima which all correspond to separate hard sub-collisions.
Besides this correlation due to sub-collisions in a single chain, we
expect that the number of chains is larger in central and smaller in
peripheral collisions. As in the models mentioned above, we want to
describe this by an impact parameter dependence, and in this paper we
have studied two different distributions, one purely stochastic and
one similar to the double Gaussian distribution used in the \pythia
model.

The third problem, concerning colour connections, is very important
for the final distribution of hadrons, but less important for a
determination of the partonic state before its hadronization. In this
preliminary study of the model we focus on features of jet and minijet
production, such as the inclusive jet production, minijet
distributions and the jet pedestal effect, which are less sensitive to
hadronization effects. We will postpone a more detailed analysis of
the colour connection structures to future work. In this study we
therefore do not include any hadronization effects, and as a
consequence we are not comparing our results with experimental data,
but are satisfied by a comparison with parton level results from the
\pythia generator.

For the numerical estimates we are using a modified version of the
\ldcmc event generator program, and as an example we study $pp$
collisions at $\sqrt{s} = 1.8$~TeV. From the numerical results we
realize that it is very essential to consider the effects of
energy-momentum conservation. This reduces the possibility to have
many chains in a single event, and it also constrains the final state
radiation in the regions close to the proton remnants.

On a qualitative level the results from our two versions agree well
with the ones obtained from \pythia. We find that our version with
uncorrelated chains gives larger fluctuations than the corresponding
model with uncorrelated scatterings in \pythia, which is expected
since the dipole chains already include some correlations between
multiple scatterings in the same chain. Including the correlations
from our geometrical chain distribution increases the fluctuations,
but not fully to the level of the model based on a double Gaussian
impact parameter dependence in \pythia. This is because the
suppression of large chain-multiplicities due to the energy
conservation procedure, is larger than the corresponding suppression
in \pythia.

As discussed above, there are still unsolved problems in the LDC model
for hadronic collisions. This concerns in particular the modeling of
colour connections between the chains and the hadron remnants in a
single event, which needs to be analysed in much more detail. In turn,
this will affect the way final state bremsstrahlung and hadronization
are handled, and also the impact of energy conservation. In
particular, the simplified model presented here probably overestimates
the soft final state emissions, since the final state radiation is
allowed from each chain independently.  With proper treatment of
colour connections we expect a reduction of such soft emissions.

Furthermore, the implementation of the model in the \ldcmc event
generator needs to be improved substantially, as the current
procedure, where $\gamma^*$-p collision events need to be reweighted
and combined by hand into p-p events, is clearly too cumbersome.

Despite the shortcomings of the simplified model presented here, we
are confident that the problems can be solved and that the final model
will be able to describe essential features of high energy hadronic
collisions, including underlying events and jet-pedestal effects in
high \ET\ events.

\section*{Acknowledgments}

We would like to thank Torbjörn Sjöstrand for valuable discussions.

\bibliographystyle{utcaps} \bibliography{references}

\end{document}